\documentclass[english,aps,pra,reprint,noshowpacs,superscriptaddress]{revtex4-1}   
\usepackage[T1]{fontenc}	% should generally be included for better accented-word behavior
\usepackage[latin9]{inputenc}	% should generally be included for better accent behavior
\usepackage{geometry}		% for controlling page margins
\usepackage{graphicx}
\usepackage[above,below]{placeins}	% allows use of \FloatBar­rier command to force section barriers
\usepackage{times}
\usepackage{hyperref}  
\usepackage{dramatist}
\hypersetup{colorlinks=true,urlcolor=blue,citecolor=blue,linkcolor=blue}   
\graphicspath{ {images/} }
\begin{document}

\title{Recurring questions that sustain the sensemaking frame}
\author{Tor Ole B. Odden}
\affiliation{Center for Computing in Science Education, Department of Physics, University of Oslo, 0316 Oslo, Norway}
\author{Rosemary S. Russ}
\affiliation{Department of Curriculum and Instruction, University of Wisconsin-Madison, Madison, WI 53706, USA}

%\keywords{sensemaking, framing, questioning}

\begin{abstract}
Many physics instructors aim to support student sensemaking in their classrooms. However, this can be challenging since instances of sensemaking tend to be short-lived, with students often defaulting to approaches based on answer-making or rote mathematical manipulation. In this study, we present evidence that specific recurring questions can serve a key role in the sensemaking process. Using a case-study of two students discussing an E\&M thought experiment, we show how students' entry into sensemaking is marked by the articulation of a particular question, based on a perceived gap or inconsistency in understanding. and how this question recurs throughout their subsequent explanations, arguing that these recurrences may serve to stabilize and extend the process.
\end{abstract}

\maketitle

\section{Introduction}
Learning physics is, in many ways, a process of learning to make sense of the world. For this reason, many physics instructors prioritize sensemaking in their teaching, aiming to support it through their curricular structure~\cite{Potter}, classroom norms~\cite{Turpen}, and through specific, sensemaking-focused activities~\cite{Gire}. However, eliciting and sustaining sensemaking is no simple task, as students may often default to a plug-and-chug~\cite{Tuminaro} or answer-making~\cite{Chen} approach to learning physics, even when learning environments are geared towards sensemaking.

Our current project is focused on defining and characterizing physics students' sensemaking processes in introductory physics. By doing so, we hope understand the factors that motivate sensemaking and how to support students in that process once they have begun. In this article, we present evidence for the role of a specific kind of student question in facilitating and sustaining the sensemaking process.

\section{Theoretical Framework}

In this study, we are drawing on the construct of \emph{framing}, a term borrowed from the sociology, linguistics, and psychology literature, referring to how individuals or groups of people answer the question "what's going on here?"~\cite{Scherr,Tannen}. Based on this construct, along with previous studies of sensemaking from PER and science education~\cite{Hutchison,Ford,Rosenberg,Odden}, we view sensemaking as a particular frame in which students aim to "figure something out"---to ascertain the mechanism underlying a phenomenon in order to resolve a gap or inconsistency in their understanding. When sensemaking, physics students spend most of their time iteratively building and revising an explanation, often out of a mix of everyday and academic knowledge. 

A key aspect of framing theory is that people shift in and out of frames, often quite quickly (on the order of a few minutes). This means that, generally speaking, we would not expect students to maintain a sensemaking frame for a long period of time. Rather, we would expect them to frequently shift between a sensemaking frame and other frames over the course of a single class period. This tendency has been observed in several studies of framing and sensemaking in science classrooms~\cite{Scherr,Hutchison,Rosenberg}. However, for instructors who are trying to prioritize sensemaking in their classrooms, this instability in the sensemaking frame can be a problem since they may wish students to sustain sensemaking discussions for longer than a few minutes.

So, what can be done to sustain the sensemaking frame? In this paper, we propose that certain types of questions can serve a key role in stabilizing sensemaking. Using a case study of two students working through a prompt in a clinical interview setting, we show how the students identify a specific inconsistency in their own understandings and express this inconsistency as a recurring question. Based on this case, we argue that these types of questions can both mark the beginning of the sensemaking process and help stabilize it once it has begun.

\section{Methods}

The data for this research comes from a study on sensemaking in introductory physics, in which we aimed to collect multiple episodes of physics sensemaking across several groups of students over the course of a semester. Our study took place with students from an introductory, algebra-based E\&M course at a major Midwestern university.

We chose to do semi-structured cognitive, clinical interviews with students from this course~\cite{Russ}, in order to try to prime them into a sensemaking frame and document the strategies they used in their sensemaking. Based on recommendations from the sensemaking literature~\cite{Ford}, we chose to interview students in pairs, in order to encourage students to check or critique each other's explanations. Approximately half of the interview questions were thought experiments that required students to discuss a hypothetical scenario involving some E\&M-related phenomena and come to a decision, usually about the safety of a particular action. For example, the case we present in this article happened in response to the following prompt: 

\begin{quote}
\emph{During a thunderstorm, you and a friend wisely decide to take shelter in your car, which you've parked in an open-air parking lot. As you are waiting out the storm, lightning strikes the car. However, besides being a little bit shaken up by the loud noise and bright flash, you both feel totally fine. After the storm has passed, you feel like getting out to stretch your legs, but as you reach for the door to get out your friend yells "Stop!" and warns you that leaving the car might be dangerous. Do you believe your friend? Why or why not?}
\end{quote}

Interviews were video and audio-recorded, transcribed, and analyzed by the first author (with input from the second author) with a particular focus on student dialogue and framing behaviors. In our analysis, we used a modified version of the categories of student framing in clinical interview settings proposed by Russ et al.~\cite{Russ}:

\begin{enumerate}

\item \textbf{Oral examination:} in this frame, the students see their task as producing a correct answer to a prompt or question in a clear and concise fashion 

\item \textbf{Expert interview:} in this frame, the students see their task as discussing their own thinking or prior knowledge on a subject, positioning themselves as an authority on that subject

\item \textbf{Brainstorming:} in this frame, students are remembering or "dredging up" their initial ideas on a subject in response to a posed prompt or question. This is similar to what Sherin et al. refer to as "mode skimming"~\cite{Sherin}

\item \textbf{Sensemaking:} in this frame, students are trying to collaboratively build a new explanation in response to a perceived gap or inconsistency in their own understandings

\end{enumerate}

After analyzing these interviews according to these framing dimensions, we began to notice that there was a critical, transitional moment in students' framing, in which they moved from the \emph{expert interview,} \emph{oral examination}, or \emph{brainstorming} frames to \emph{sensemaking}. This transition seemed to be accompanied by a verbalized question or statement of uncertainty, and students often returned to this same question multiple times throughout their episodes of sensemaking. Once we had identified this transition, we specifically focused our analysis on the role that these particular questions or statements of uncertainty played in the process of sensemaking.

In the next section, we present a case study from this data corpus illustrating the role that one of these questions played in the explanation that two students generated in response to the aforementioned lightning safety prompt. Case studies are useful for developing plausible existence arguments and to identify the underlying mechanisms for behavioral phenomena ~\cite{Lising}, so we feel this presentation is appropriate for this type of analysis. The students, Jake and Liam, were friends who were concurrently enrolled in the same course.

\section{A Case of Repeated Questioning in Sensemaking}

After drawing a sketch of the car/lightning system as shown in Figure 1, Jake and Liam initially reasoned that after the lightning strike the charge will spread out throughout the car and will then gather in the car's "pointy parts," in the same way that charge accumulates in static discharge wicks on airplane wings. Focusing specifically on the door handle mentioned in the prompt, they reasoned that because the handle isn't one of these "pointy parts" the person inside the car would at least be safe in grabbing that handle. However, this reasoning led them to a key question: how does charge ever leave the car?

\begin{figure}
\includegraphics[width=0.8\linewidth]{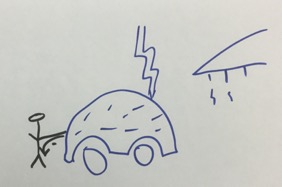}
\caption{Jake and Liam's drawing of the "lightning car" scenario, with the analogous airplane wing on the right side\label{fig1}}
\end{figure}

\begin{drama}
  \Character{J}{J}
  \Character{L}{L}
  \Jspeaks: Oh, okay, so, conductor---uh, this is a really, like, it's a really big surface, so the electrons are gonna like spread out across, like this. And, because there's like, like the door handle of our car isn't like a point, it's not gonna like build up right there, where like a shock's gonna come off of it. Because remember, like the airplane wings has these like little, like, points there and that's were like the shocks come off. So there's nowhere on the car that's like a big point like that. So, it's not gonna shock you when you reach for the door handle because there's not enough, like, electrical, electr---uh, like a negative electric charge that, uh, create a shock. (silence, 14s)
\Lspeaks: I think that makes sense.
\Jspeaks: \textbf{I just don't know how you get rid of this [charge].}
\Lspeaks: Yeah, like does your car just stay charged until it gets grounded?
\Jspeaks: Yeah, 'cause I al---I know, like, like the, I feel like the 
severe threat would be like your, like, your radio and all that stuff on the inside, like that might be like affected by all this like negative charge coming in. But, like, we're not gonna get shocked reaching for a door handle.
\end{drama}

In terms of the framing codes used in the analysis, we argue that during this exchange, the pair appears to have shifted from {\em brainstorming} to {\em sensemaking}. Initially, they were focused on "dredging up" knowledge relevant to the interview prompt, such as the fact that charge on airplane wings tends to accumulate in points. During this period, they are focused on satisfying the prompt, and do not appear to have noticed any particular gap or inconsistency in their own understandings. Once they have laid out these initial ideas, however, Jake notices an inconsistency in the scenario they've described: it seems intuitive that the car should become de-charged at some point, but in the described scenario Jake wasn't able to see a mechanism for how that would happen. Jake points out this inconsistency out with the question "how does the charge leave the car?" and the pair begin to try to "figure out" this question by iteratively building and revising an explanation to resolve it.

The discussion continues, with the interviewer bringing the topic back to the safety of actually leaving the car rather than just reaching for the door handle: 

\begin{drama}
 \Character{J}{J}
 \Character{L}{L}
\Character{I}{I}
 \Ispeaks: So, I've heard some folks say that the real problem is [...] once you step out, you put a foot on the ground, and you're getting out of the car, that's where the danger would be.
\Lspeaks: I can see how that would [be dangerous]---'cause at that point you're the grounding mechanism. 'Cause charge wants to go down, like it wants to get out of there. So it's like, when you touch that it's all just gonna rush down through you, like all the negative charge.
\Jspeaks: Yeah. So I don't know how you would---but \textbf{I don't know how you'd get rid of this charge, then.} I don't know if it just, like, as you drive, it just, like... *sigh* (silence, 12s) 'Cause I'm pretty sure lightning has a pretty good negative charge to it, where the fact that it's gonna deposit a large amount of current to like shock you
\Lspeaks: Yeah. (silence, 9s) I dunno, I guess you don't ever hear about cars being struck where they have to, like, de-charge the car.
\end{drama}

In this segment the pair remain in a sensemaking frame, as they continue to try to build their explanation. Additionally, we see Jake return to his question and the underlying inconsistency: how do you get rid of the charge on the car, if there isn't any direct path between the car body and the ground? This recurrence of the question is phrased almost identically to his first articulation.

At this point, having seen how the pair's explanation has developed, we can begin to unpack the role of Jake's question in the pair's sensemaking conversation. Jake's initial question, we argue, facilitated the pair's shift into a sensemaking frame by determining both the topic and goal of their conversation going forward. That is, Jake's statement "I don't know how you'd get rid of this charge" in the previous segment focused the pair's attention on a specific aspect of the physical situation they were analyzing, the free charge trapped on the surface of the car. In that way, it set the topic of their conversation. But, by pointing out an unresolved point in their explanation, the utterance also set Jake and Liam's goal for the rest of the episode: to generate an explanation for how the charge would leave the car.

What role did the first recurrence of the question serve in the conversation? Our conjecture is based on Liam's statement, "I guess you don't ever hear about cars being struck where they have to, like, de-charge the car." Until this point, the pair had been focusing their conversation on the physical situation in the abstract, discussing what hypothetical charges would do on their hypothetical car. In the next line, however, Liam takes the explanation in a different direction than before, using what does or doesn't happen in real life as a resource in the explanation. Looking at this statement in relation to Jake's restatement of the initial question, it seems that Jake may have provided an opening for that move by implying that, in theory, one should be able to get rid of the charge on the car. By implying that it is possible to get rid of this charge, this repeat of the question may have opened up new possible directions for the pair's explanation.

Jake articulated his question one more time, a few moments later in the episode:

\begin{drama}
 \Character{J}{J}
 \Character{L}{L}
\Character{I}{I}
\Lspeaks: I don't know. (silence, 6s) 
\Jspeaks: But yeah, I understand the... (silence, 8s) \emph{*sigh*} I guess, I guess my inclination would be to answer the question is you wouldn't get electrocuted just by touching it when you get out, like getting out of the car wouldn't be an issue, but, like, I mean like if you were to touch the frame of the car, once you get outside of it, I don't think that you'd get electrocuted because... \emph{*sigh*} And I don't know how this mechanism happens, but I just feel like that there's a way where like the car gets struck but, like, the charge doesn't stay on there that long. So unless you are like touching the car when it gets struck then obviously you'd get electrocuted but there's a mechanism, \textbf{I don't know what it is,} [...] \textbf{I just don't know the mechanism by where this charge goes---I don't know where it goes.}
\end{drama}

At this point, sensing some growing frustration, the interviewer decided to throw out a hint to the pair, suggesting that this situation may be less dangerous in thunderstorms, and the pair quickly settled on the explanation that conductive rain will carry the charge off of the car.

During this segment, we argue that Jake was still in a sensemaking frame---that is, he was still trying to "figure out" an explanation to resolve his perceived inconsistency in knowledge. And, looking at Jake's utterance in this segment, we see him returning to his question one more time, saying "I just don't know the mechanism by where this charge goes."
 
What role does this second recurrence serve in the conversation? Based on Jake's repeated emphasis of his uncertainty (saying "I don't know" four times in the span of a single minute) we would argue that he seems to be using the question to express his dissatisfaction with the explanation they've generated. This happened shortly after two long silences, which seemed to indicate that the pair were running dry on ideas for how to resolve this inconsistency. So, without anywhere else to go in his explanation, Jake seems to be reiterating that there should be a mechanism here, he just cannot see it. Unlike the previous recurrence, this statement did not necessarily open up new avenues for the conversation; rather, by expressing dissatisfaction with the explanation so far, it seems to be a bid to simply keep their sensemaking going---to continue moving towards an explanation by any means possible.

In summary, in this case we see Jake pose the same question on three different occasions during a sensemaking conversation. Based on the conversational cues, we are arguing that the first instance of the question served to focus the pair's attention on a particular gap in their understanding; the second instance opened up new avenues for their explanation; and the third instance allowed Jake to express his dissatisfaction with the explanation they had generated.

\section{Discussion and Conclusion}

Science education researchers have, for decades, argued that question-asking is a key aspect of student inquiry~\cite{Chin}. By asking deep, meaningful questions, students are able to uncover and then resolve gaps and inconsistencies in their understanding, thereby strengthening their knowledge frameworks. In this case, we see just such a process at work. Although the questions were formulated slightly differently each time, we argue that they all are referring back to the same perceived inconsistency: a missing mechanism for how charge could safely leave the car. The overall process of building an explanation to resolve this inconsistency is, by our definition, sensemaking, and so one major takeaway from this case study is that these types of recurring questions may well be characteristics of the sensemaking process. 

However, we are going one step further and arguing that these types of questions may be more than just characteristics of sensemaking: they can also serve particular conversational functions, helping to facilitate and stabilize the sensemaking frame. In this case, the initial instance of the question marked the students' transition into a sensemaking frame, focusing their attention on a particular unresolved issue in their initial explanation. Thereafter, we suggest that the recurrences helped to keep sensemaking going by opening up new directions in the conversation and allowing Jake to express his frustration with the lack of resolution. By repeatedly returning to this question Jake was then able to move the conversation forward, and this progress, we argue, helped to maintain the frame. Thus, we are proposing that these kinds of recurring questions may help to counteract the inherent "slipperiness" or dynamism of frames like sensemaking, keeping students engaged when they otherwise might move on. Of course, there are other factors at work as well; in this case, a specific nudge from the interviewer also helped to keep the students focused on Jake's question; however, in the absence of this kind of vexing inconsistency in knowledge we are doubtful that such a nudge would have been nearly as effective. 

If, indeed, such repeated questions do serve a key role in sensemaking, we feel that they could be a valuable tool for sensemaking-focused physics education research and/or instruction. For physics education researchers these types of recurring questions could give us a useful marker for analytically identifying sensemaking in qualitative data. That is, if students are repeatedly returning to a vexing question while they are trying to build an explanation (either individually or in a group setting), that could serve as evidence that those students are in a sensemaking frame. Practically, teachers aiming to support sensemaking may want to leverage these kinds of questions in the classroom. For example, one could keep track of which specific questions drive students towards this kind of explanation-building and try to use them as discussion prompts, and/or set the learning environment up so that students are encouraged to elicit and follow up on these types of questions. In either case, we hope that by attending to these types of questions, instructors and researchers may have the opportunity to see many more of these rich sensemaking discussions.

% For short, simple bibliographies, manually formatting works:
% Remember that you'll need to run pdfLaTeX twice to get the references to show (first 
% pass will insert ?? in their places).

% For a longer bibliography, delete the thebibliography block above, then comment in 
% these two lines to use a .bib file with BibTeX.
%\bibliographystyle{apsrev}  	% supercedes the longbibliography option, so leave commented out if you want to display article titles
%\bibliography{mybibfile}  	% don't include the .bib suffix

\end{document}